\documentclass[twocolumn,11pt]{article}
\usepackage[margin=0.75in]{geometry}
\usepackage{longtable}
\usepackage{booktabs}
\usepackage{array}
\usepackage{pdflscape}
\usepackage{ragged2e}
\usepackage[utf8]{inputenc}
\usepackage[T1]{fontenc}
\usepackage[lining,tabular]{ebgaramond}
\usepackage{amsmath}
\usepackage{amssymb}
\usepackage[cmintegrals,cmbraces]{newtxmath}
\usepackage[final,babel=true]{microtype}
\usepackage{hyperref}
\usepackage{enumitem}
\usepackage{graphicx}
\usepackage{algorithm}
\usepackage{algpseudocode}
\usepackage{tikz}
\usepackage[backend=biber, style=authoryear-comp, citestyle=authoryear]{biblatex}
\usetikzlibrary{shapes,arrows.meta,positioning,fit,backgrounds,calc,decorations.pathreplacing}
\usepackage{lettrine} 
\usepackage{titlesec}
\titleformat{\section}
  {\normalfont\large\scshape\centering}{\thesection}{1em}{}
\titleformat{\subsection}
  {\normalfont\normalsize\itshape}{\thesubsection}{1em}{}

\newcolumntype{L}[1]{>{\RaggedRight\arraybackslash}p{#1}}

\newcommand{\Dasein}{\textit{Dasein}}
\newcommand{\Entwurf}{\textit{Entwurf}}
\newcommand{\Gestell}{\textit{Ge-stell}}
\newcommand{\Bestand}{\textit{Bestand}}
\newcommand{\DasMan}{\textit{Das Man}}
\newcommand{\Einebnung}{\textit{Einebnung}}
\newcommand{\Eigentlichkeit}{\textit{Eigentlichkeit}}
\newcommand{\Gerede}{\textit{Gerede}}
\newcommand{\Lichtung}{\textit{Lichtung}}
\newcommand{\Herausfordern}{\textit{Herausfordern}}
\newcommand{\Gelassenheit}{\textit{Gelassenheit}}
\newcommand{\Zuhanden}{\textit{Zuhanden}}
\newcommand{\Vorhanden}{\textit{Vorhanden}}
\newcommand{\Sorge}{\textit{Sorge}}

\newcommand{\Angst}{\textit{Angst}}
\newcommand{\Zuhandenheit}{\textit{Zuhandenheit}}

\newcommand{\Aletheia}{\textit{Aletheia}}
\newcommand{\SeinzumTode}{\textit{Sein-zum-Tode}}
\newcommand{\Gewesenheit}{\textit{Gewesenheit}}
\newcommand{\Weltlosigkeit}{\textit{Weltlosigkeit}}

\title{\textbf{The Metaphysics We Train: A Heideggerian Reading of Machine Learning}}
\author{Heman Shakeri \\ School of Data Science, University of Virginia}

\date{\vspace{-0.5em}\small\itshape Dedicated to my friend Phil Bourne, who recognized data science as a way of thinking that transcends the technological.}

\begin{document}

\maketitle

\begin{abstract}
This paper offers a phenomenological reading of contemporary machine learning through Heideggerian concepts, aimed at enriching practitioners' reflexive understanding of their own practice. We argue that this philosophical lens reveals three insights invisible to purely technical analysis. First, the algorithmic \Entwurf\ (projection) is distinctive in being automated, opaque, and emergent---a metaphysics that operates without explicit articulation or debate, crystallizing implicitly through gradient descent rather than theoretical argument. Second, even sophisticated technical advances remain within the regime of \Gestell\ (Enframing), improving calculation without questioning the primacy of calculation itself. Third, AI's lack of existential structure, specifically the absence of Care (\Sorge), is genuinely explanatory: it illuminates why AI systems have no internal resources for questioning their own optimization imperatives, and why they optimize without the anxiety (\Angst) that signals, in human agents, the friction between calculative absorption and authentic existence. We conclude by exploring the constructive implications of this diagnosis: relocating AI from defective \Dasein-simulacrum to transparent \Zuhanden\ (ready-to-hand) equipment, and arguing that data science education should cultivate not only technical competence but ontological literacy---the capacity to recognize what worldviews our tools enact and when calculation itself may be the wrong mode of engagement.
\end{abstract}

\section{Introduction: A Philosophical Lens on Technical Practice}

When we train a neural network, what exactly are we doing? The standard account is epistemological: we are discovering patterns latent in data. Yet a phenomenological reading suggests something quite different---we are enacting an ontological projection, a mathematical framework that determines what \textit{can} appear as knowable.

This paper reads contemporary machine learning through Martin Heidegger's phenomenology \parencite{heidegger1962being, heidegger1977question, heidegger1967thing}. Our aim is not to identify failures or prescribe solutions, but to use continental philosophy as a lens that makes visible the implicit ontological commitments embedded in our computational tools.

We are particularly interested in three insights that emerge from this reading:

\begin{enumerate}[leftmargin=*]
    \item \textbf{The Distinctiveness of Algorithmic Projection:} While all science involves theoretical frameworks (Newton's spatiotemporal manifold, Darwin's evolutionary tree), the algorithmic \Entwurf\ exhibits unique characteristics---it is automated, opaque, high-dimensional, and emergent. Unlike Newtonian mechanics, which can be explicitly debated, the metaphysics of a trained neural network operates implicitly, revealed only through its effects.
    
    \item \textbf{Why Sophistication Does Not Escape \Gestell:} We argue that modern advances represent a paradox: we have increased the sophistication (ontic complexity) without altering the \Gestell\ (ontological essence). Even technically advanced approaches---Bayesian deep learning, fairness methods, uncertainty quantification---remain within the regime of Enframing. They improve calculation without questioning the primacy of calculation itself.
    
    \item \textbf{The Existential Deficit as Genuinely Explanatory:} AI's lack of existential structure is not philosophical flourish but explanatory. Without Care (\Sorge), AI systems are trapped within their optimization imperatives. They cannot question whether optimization is appropriate, recognize category errors, or challenge their own goals. This explains certain fundamental limitations better than purely technical accounts.
\end{enumerate}

\subsection{An Interpretive Mapping, Not an Identity}

This is an \textit{interpretive} exercise. Crucially, we do not claim a strict ontological identity between machine operations and human existential structures. We do not assert that a softmax function ``is'' exactly \DasMan, or that a neural network possesses \Dasein\ or experiences anxiety. In Heidegger's terms from his 1929--1930 lectures, while the human is world-forming (\textit{weltbildend}), the machine is fundamentally worldless (\textit{weltlos}) \parencite{heidegger1995fundamental}.

Instead, we deploy phenomenological vocabulary as an interpretive mapping---a bird's-eye hermeneutic lens that helps us make sense of ML concepts and navigate them better. We read ML architectures, loss functions, and training procedures as objects worthy of philosophical attention---much as one might read a novel, a painting, or a social practice. By asking how operations like multi-head attention or generative diffusion \textit{structurally parallel} concepts like \Gestell\ or \Bestand, we gain a vocabulary to articulate the implicit ontological commitments embedded in our computational tools. The value lies in achieving an enriched, reflexive understanding of our engineering practice, not in proving a literal equivalence.

Technical solutions to the problems we identify exist---indeed, we will acknowledge them. But the philosophical lens shows that the \textit{kind} of problem runs deeper than choosing better architectures or loss functions. It concerns the worldview our tools enact, often without our awareness.

\subsection{Situating This Work: From Dreyfus to Deep Learning}

Our approach builds on a long tradition of phenomenological critique, but it is not a repetition. If \textcite{dreyfus1992computers} dismantled the ``Old Testament'' of AI---Symbolic AI, or what he called Good Old-Fashioned AI (GOFAI)---this paper offers a critique of its ``New Testament'': connectionism, deep learning, and the Transformer architecture.

Dreyfus's target was the \textit{Psychological Assumption}: the belief that human intelligence operates by following formal rules on discrete symbols. He argued that human understanding relies on a background of embodied common sense and ``fringe consciousness'' that cannot be formalized into explicit rules. The mind, he insisted, does not ``count out'' alternatives; it ``zeros in'' on relevance through a global, situated grasp of context that rule-based systems could never achieve.

Modern AI practitioners often dismiss Dreyfus on precisely these grounds: ``We don't use rules anymore; we use neural networks that learn from data.'' Transformers, in particular, appear to have solved the mechanical problem Dreyfus identified. The attention mechanism can ``zero in'' on relevant features without an infinite regress of context-specifying rules. Does this not vindicate connectionism against the Heideggerian critique?

Our argument is that it does not. While deep learning escapes the rule-based trap Dreyfus identified, it falls into a different one: \textit{optimization as ontology}. Replacing explicit rules with high-dimensional vector spaces does not resolve the Heideggerian problem; it relocates \Gestell\ from logic to geometry. Where Dreyfus critiqued symbolic AI's inability to capture context, we critique the \textit{way} connectionist AI captures context---by reducing it to calculable proximity in embedding space, by mathematically \textit{demanding} relevance through dot products rather than \textit{caring} about it.

This is why our reading is scientifically coherent despite---indeed, \textit{because of}---modern AI's engineering successes. Dreyfus was often skeptical that the engineering would ever work; he doubted computers would beat chess champions or handle ambiguity. History proved him partially wrong on the engineering. But we grant the engineering success precisely to reveal the ontological emptiness it conceals. The question is no longer ``Can the machine perform the task?'' but ``What is the machine doing when it performs the task?''

Here we introduce a distinction that updates Dreyfus for the probabilistic era: modern AI handles \textit{uncertainty}---a mathematical quantity amenable to probability distributions---but not \textit{ambiguity}---an existential condition requiring judgment, stakes, and Care. A Bayesian model can represent that it is 60\% confident in one interpretation and 40\% in another. But it cannot recognize that some situations demand \textit{commitment} rather than probability, that certain choices are incommensurable, that occasionally the right response is not to optimize at all. Probabilistic sophistication is not existential depth.

Our specific contributions---the analysis of Softmax as structurally paralleling the ``publicness'' of \DasMan, the reading of attention as \Herausfordern, the link between ERM and the absence of \Sorge---are critiques Dreyfus could not have made because the mathematical structures did not yet dominate AI. We identify how the \textit{specific mathematics} of 2025---logits, softmax normalization, gradient-based loss minimization---enact particular ontological commitments. The apparatus has changed; the \Gestell\ persists.

More recent work in perspectivism \parencite{giere2019scientific} and practice-based philosophy of science \parencite{rouse2002scientific} has similarly stressed that scientific models project partial, practice-laden worlds. \textcite{winograd1986understanding} account of software as ontological design remains foundational. Our contribution is to bring these insights to bear on the specific architectures and training paradigms of contemporary deep learning, showing \textit{what particular commitments} emerge from attention mechanisms, normalization layers, and gradient-based optimization---and why increased sophistication does not constitute escape from Enframing.

\section{The Mathematical Projection in Science: From Newton to Neural Networks}

\subsection{Heidegger's Critique of the Mathematical}

To understand what makes algorithmic projection distinctive, we must first grasp Heidegger's analysis of modern science. In \textit{What Is a Thing?} \parencite{heidegger1967thing}, Heidegger argues that the essence of the ``mathematical'' is not calculation but \textit{projection} (\Entwurf)---the establishment of an \textit{a priori} framework that determines what can count as a phenomenon.

Before Newton could measure gravitational force, he first had to project nature as a spatiotemporal manifold governed by mathematical law. This projection is not \textit{derived from} observation; it makes observation possible. As \textcite{hanson1958patterns} argued, observation is theory-laden---we see what our frameworks allow us to see.

In contemporary machine learning, the ``Deep Learning Workflow'' similarly projects nature as a probabilistic latent manifold accessible through differentiable optimization. The question is: \textit{what makes this projection distinctive?}

\subsection{The Distinctiveness of the Algorithmic \textit{Entwurf}}

The algorithmic projection differs from previous scientific frameworks in four critical dimensions:

\subsubsection{Automation and Emergence}
Newtonian mechanics was explicitly formulated, debated, refined. Scientists could articulate and critique its assumptions. In contrast, a trained neural network's ``theory of the world'' is \textit{emergent}---it crystallizes implicitly through gradient descent over millions of parameter updates. No human specifies the resulting worldview; it emerges automatically from data and architecture.

As \textcite{olah2020zoom} observe, features in deep networks are not designed but discovered during training. We cannot point to a line of code that says ``this is how the model understands faces.'' The projection emerges, operates, and shapes downstream decisions---all without explicit articulation.

\subsubsection{Opacity and Illegibility}
Newton's $F = ma$ is inspectable. One can debate whether force is ontologically primitive or whether mass requires clarification. The projection is \textit{legible}.

A trained Transformer's 175-billion-parameter projection is fundamentally opaque \parencite{lipton2018mythos}. Even mathematically elaborate interpretability methods (attention visualization, feature attribution) provide only fragmentary glimpses \parencite{rudin2019stop}. The metaphysics operates in a high-dimensional space inaccessible to human intuition.

\subsubsection{High-Dimensional Realization}
Classical scientific theories operate in human-intelligible dimensionality. The algorithmic projection realizes itself in spaces of millions or billions of dimensions. The very structure of how it organizes reality exceeds our cognitive grasp.

This is not merely a practical limitation. It represents a qualitative shift: a worldview that cannot, even in principle, be fully articulated in human language or visualized in human-perceptible space.

\subsubsection{Performative Deployment}
Newtonian mechanics describes; neural networks \textit{enact}. When a recommender system is deployed, it doesn't merely predict behavior---it shapes it. The projection becomes performative, creating feedback loops where the model's worldview actively constructs the reality it purports to describe \parencite{mackenzie2006engine}.

This is the transition from theoretical framework to what \textcite{heidegger1977question} termed \Gestell---a mode of revealing that actively orders the world according to its own logic.

\subsection{Implications of Distinctiveness}

These characteristics mean the algorithmic \Entwurf\ operates as a kind of \textit{infrastructural metaphysics}---shaping what appears as real, valuable, or possible without ever being explicitly stated or debated. As \textcite{bowker2000sorting} argue about classification systems, it becomes ``infrastructure''---invisible, taken-for-granted, yet profoundly consequential.

This is not to say algorithmic projection is necessarily harmful. It is to say it requires a different mode of critical engagement than previous scientific frameworks. We cannot simply ``read the theory'' because the theory is never written. We must attend to its effects, its structural tendencies, its implicit commitments.

\section{Inductive Bias as Ontological Commitment}

\subsection{The Phenomenology of Architectural Choice}

When a practitioner selects a specific neural architecture, they are not merely choosing a computational tool. They are selecting what Heidegger would call a mode of ``taking-as''---a framework that determines how phenomena can appear.

In machine learning, this is formalized as \textit{inductive bias}: the set of assumptions a model makes about the structure of the learning problem \parencite{mitchell1980need}. An architecture designed for images (CNNs) cannot easily learn sequential dependencies; one designed for sequences (RNNs) struggles with spatial hierarchies.

From a phenomenological perspective, inductive bias is \textit{ontological design}. As \textcite{winograd1986understanding} argued, ``in designing tools we are designing ways of being.'' The architecture projects a world-structure; training merely finds the specific parameters within that structure.

\subsection{The Topology of Revelation and Concealment}

Every projection reveals by concealing. A CNN that learns hierarchical visual features excels at object recognition but loses temporal context. An attention mechanism that captures long-range dependencies may struggle with fine-grained local structure.

This is not a bug but an ontological necessity. Heidegger's concept of \textit{aletheia} (unconcealment) emphasizes that truth is always perspectival---to reveal is simultaneously to conceal \parencite{heidegger1962being}. The architecture establishes a ``clearing'' (\Lichtung) where certain patterns appear, necessarily leaving others in shadow.

Technically, this manifests in the phenomenon of \textit{negative transfer} in multi-task learning \parencite{standley2020tasks}. When a single backbone is forced to optimize for ontologically distinct tasks, gradients conflict. The network attempts to inhabit two incompatible clearings simultaneously, resulting in what we might call \textit{ontological incoherence}.

\section{The Transformer as \textit{Ge-stell}: A Case Study}

To make the phenomenological reading concrete, we examine the Transformer architecture \parencite{vaswani2017attention} as a paradigmatic instantiation of \Gestell---Heidegger's concept of ``Enframing,'' the essence of modern technology that reveals beings as Standing-Reserve (\Bestand), resources optimized for ordering and control.

\subsection{Embedding Space as Mathematical Clearing (\textit{Lichtung})}

The Transformer begins by projecting discrete tokens into a continuous embedding space $\mathbb{R}^{d}$. This space functions as what Heidegger called the ``clearing''---a region where meaning can appear.

For the algorithm, this latent space \textit{is} the world. Semantic relationships manifest as geometric proximities; similarity becomes Euclidean distance; meaning becomes vector operations. The embedding space is not a representation \textit{of} some external semantic reality---it is the horizon within which meaning can appear at all.

Crucially, this revealing is also a concealing. By projecting meaning into continuous, high-dimensional Euclidean geometry, alternative ontologies---discrete symbol manipulation, structured logic, embodied grounding---are not eliminated but strongly constrained. The space has a topology, and that topology determines what meanings can be proximate, what transformations are continuous, what patterns are learnable.

\subsection{Attention as Challenging-Forth (\textit{Herausfordern})}

The attention mechanism performs:
\begin{equation}
\text{Attention}(Q,K,V) = \text{softmax}\left(\frac{QK^T}{\sqrt{d_k}}\right)V
\end{equation}

Through our interpretive mapping, this mechanism structurally parallels what Heidegger termed ``challenging-forth'' (\Herausfordern)---the active demanding that nature present itself in calculable form \parencite{heidegger1977question}.

The Query vector ($Q$) actively interrogates the sequence (Keys, $K$), \textit{demanding relevance} through the dot product $Q \cdot K$. This is not passive observation but mathematical demanding-to-appear. The mechanism doesn't ask ``what is naturally related?'' but rather ``\textit{become} related to me according to this metric.''

The Values ($V$) are then aggregated according to these demanded relevances, weighted by the softmax-normalized attention scores. Information is extracted, ordered, made available---\textit{challenged forth} into a calculable aggregate.

\subsection{Multi-Head Attention as Pluralized Clearings Within \textit{Ge-stell}}

Modern Transformers do not rely on a single attention mechanism, but on Multi-Head Attention (MHA). Mathematically, each head $h$ induces a trainable bilinear similarity metric $M_h := (W_Q^{(h)})^\top W_K^{(h)}$. Each head specifies a distinct, learned criterion of relevance, gathering a specific context vector $y_i^{(h)}$ for position $i$.

Through our interpretive mapping, MHA implements a pluralized disclosure. Each head acts as a partial clearing (\Lichtung), enacting a distinct ``as-structure.'' For instance, in the ``world of baseball,'' a sentence about a ``closer,'' a ``ground ball,'' and a ``double play'' opens a determinate world of stakes, roles, and physical actions. Within a Transformer, different heads approximate this by binding role terms, action-object links, and narrative salience in parallel.

However, this plurality does not yield a lived \Zuhanden\ (ready-to-hand) worldhood. The model's relation to the tokens remains fundamentally calculative. MHA simply multiplies the enframing: it pluralizes the ``ways of mattering''---running multiple \Gestell\ frameworks in parallel---only to recompose them via a projection matrix $W_O$ into a single commensurable vector space. It is a highly sophisticated calculus of proximity, not an escape from calculation.

\subsection{Softmax and the Enforced Commensurability on the Simplex}

The softmax normalization deserves particular attention:
\begin{equation}
\alpha_{ij} = \frac{\exp(s_{ij})}{\sum_{j} \exp(s_j)}
\end{equation}

This operation exhibits three structural characteristics suggestive of what Heidegger called the ``publicness'' (\textit{\"Offentlichkeit}) of \DasMan\ (``The They''):\footnote{Throughout, we use Heideggerian terms such as \Einebnung\ and \DasMan\ analogically to illuminate structural features of contemporary ML practice. We do not claim that operations like softmax or MSE ``are'' these existential structures in any strict sense, but that the patterns they instantiate in common socio-technical configurations can be usefully read through this phenomenological vocabulary.}

\subsubsection{Enforced Relativization}
Softmax makes all scores \textit{relative} rather than absolute. A token's relevance is always measured against all others in the sequence. Through our lens, this mirrors the relativistic mode of understanding characteristic of \DasMan, where nothing has intrinsic significance---everything is evaluated comparatively within the public horizon.

\subsubsection{Zero-Sum Commensurability}
Because $\sum_j \alpha_{ij} = 1$, attention is zero-sum. Attending more to one position requires attending less to others. This enforces a competitive scarcity regime and, more fundamentally, renders all positions commensurable---directly comparable on a single scale.

In Heideggerian terms, this echoes the ``publicness'' where everything must be rendered publicly comparable to count. Incommensurable forms of significance---what \textit{cannot} be compared---are structurally excluded.

\subsubsection{Homogenization Through Normalization}
Softmax's exponential nature amplifies relative differences in the scores, but because it always produces a normalized distribution with non-zero mass on every position, all candidates are forced into a single commensurable space. Even extreme preferences remain differences of degree within the same probability simplex. The result is not smoothing in the sense of equalizing scores, but \textit{homogenization} in the sense that all forms of significance must appear as comparable weights on a shared normalized scale.

Research on attention mechanisms reveals this empirically: pure attention networks can exhibit ``rank collapse,'' where representations become increasingly similar \parencite{dong2021attention}. The repeated application of this enforced commensurability across layers contributes to the gradual erosion of distinctions, paralleling Heidegger's concern that \Einebnung\ (leveling) threatens the collapse of differences into undifferentiated averageness.

The engineered nature of this ontology is made empirically visible by the softmax ``temperature'' parameter ($T$) used during inference:
\begin{equation}
\alpha_{ij}(T) = \frac{\exp(s_{ij}/T)}{\sum_{j} \exp(s_j/T)}
\end{equation}
Practitioners literally turn a mathematical dial to control the ontology of the output. A high temperature ($T \to \infty$) flattens the distribution, mathematically enforcing complete \Einebnung---total averageness and uniformity. A low temperature ($T \to 0$) forces a one-hot decision, driving the system toward dogmatic certainty---pure algorithmic \Gerede\ (idle talk), where the model asserts the most probable token with absolute, artificial conviction. The entire spectrum of disclosure, from undifferentiated publicness to unreflective assertion, is parameterized by a single scalar.

\subsection{The Present-at-Hand Limitation}

Heidegger distinguished between \Zuhanden\ (ready-to-hand, absorbed skillful use) and \Vorhanden\ (present-at-hand, detached theoretical observation). A hammer ready-to-hand withdraws into transparent use; a broken hammer becomes present-at-hand as an object for contemplation.

Attention mechanisms operate exclusively in the \Vorhanden\ mode. They treat tokens as objects with measurable properties (embeddings) to be calculated over. The absorbed, contextually embedded character of language-in-use---its ready-to-hand being---cannot appear within this framework.

As \textcite{dreyfus1992computers} argued, computational systems fundamentally lack ``Being-in-the-world.'' They process disembodied representations with no genuine stakes, no care-structure (\Sorge), no lived context. The Transformer sees language as standing-reserve---resources for calculation---never as the medium of a world we inhabit.

\section{The Algorithmic Leveling: Loss Functions and \textit{Das Man}}

\subsection{The Structure of Empirical Risk Minimization}

The dominant training paradigm---Empirical Risk Minimization (ERM)---minimizes:
\begin{equation}
\hat{L}(f) = \frac{1}{n}\sum_{i=1}^{n} \ell(f(x_i), y_i)
\end{equation}

This appears mathematically neutral---simply fitting a function to data. Yet when read phenomenologically, ERM exhibits a structural pattern suggestive of what Heidegger termed \Einebnung\ (``leveling down''), the process by which \DasMan\ (``The They'') reduces all possibilities to standardized averageness \parencite{heidegger1962being}.

\subsection{Mean Squared Error and the Ontology of Averaging}

Consider the canonical loss for regression, Mean Squared Error (MSE):
\begin{equation}
L(f) = \mathbb{E}[(Y - f(X))^2]
\end{equation}

The optimal predictor under this loss is the conditional expectation:
\begin{equation}
f^*(x) = \mathbb{E}[Y \mid X = x]
\end{equation}

In unimodal settings, this is unproblematic. But consider a bimodal distribution: two Gaussian clusters with equal weight, $Y \mid X=x \sim \frac{1}{2}\mathcal{N}(\mu_1(x), \sigma^2) + \frac{1}{2}\mathcal{N}(\mu_2(x), \sigma^2)$.

The optimal predictor is $f^*(x) = (\mu_1(x) + \mu_2(x))/2$---the \textit{average} of the two modes. This lies in a low-density region representing \textit{neither} actual behavior pattern. The model predicts a ``compromise'' that no actual instance exhibits.

This is not a failure of MSE for its intended purpose---minimizing squared error. Rather, it reveals the \textit{ontology} implicit in the loss: heterogeneous possibilities should be collapsed to a single central tendency. Deviation from the mean appears only as error to be minimized.

We can read this structurally as analogous to \Einebnung: the systematic reduction of diverse, heterogeneous modes of being to a standardized average. \DasMan\ enforces averageness (\textit{Durchschnittlichkeit})---not maliciously, but as its basic mode of understanding. Similarly, MSE privileges statistical density; minority patterns appear only as residuals.

\subsection{Cross-Entropy and Artificial Certainty}

In classification, Cross-Entropy (CE) loss is standard:
\begin{equation}
L_{CE} = -\sum_{i=1}^{C} t_i \log(p_i)
\end{equation}

Combined with high-capacity models and extensive training, CE can drive the model toward extreme confidence in a single class, even for ambiguous instances.

This creates what we might call ``algorithmic \Gerede\ (idle talk)''---Heidegger's concept of discourse that circulates average understanding without genuine appropriation, closing off inquiry through assertion of certainty \parencite{heidegger1962being}.

The model ``repeats'' the anonymous statistical majority with high confidence, exhibiting no uncertainty even at decision boundaries, no awareness of its own limitations, no recognition of ambiguity. Like \Gerede, it speaks with the voice of ``The They''---not from genuine understanding but from averaging over public opinion (the training distribution).

\subsection{RLHF and the Industrialization of Publicness}

This algorithmic leveling is perfected in modern Generative AI through Reinforcement Learning from Human Feedback (RLHF). RLHF explicitly trains the model to optimize for the aggregated preferences of anonymous crowd-workers.

Structurally, RLHF enacts the algorithmic enforcement of \textit{\"Offentlichkeit} (publicness). It mathematically penalizes idiosyncratic, authentic, or abrasive edge cases to ensure the model always speaks in the polite, harmless, leveled-down average voice of the public. The model validates its outputs not against Truth (\Aletheia), but against the statistical average of human raters---the literal voice of \DasMan.

This is not a critique of RLHF's engineering aims, which are often legitimate (reducing harmful outputs, improving helpfulness). It is a phenomenological observation about the \textit{kind of entity} the process produces: a system whose every utterance is calibrated to the anonymous public, whose ``character'' is an artifact of statistical aggregation. The result is a machine that speaks fluently in the mode of \Gerede---plausible, socially normalized discourse that circulates without genuine appropriation.

\subsection{Technical Responses and Why Philosophy Still Matters}

We must acknowledge that machine learning has developed rigorous responses. Mixture Density Networks \parencite{bishop1994mixture} explicitly model multimodal distributions rather than collapsing to means. Quantile Regression predicts distributional percentiles rather than central tendencies. Label Smoothing \parencite{szegedy2016rethinking} reduces overconfidence by softening one-hot targets. Calibration methods such as temperature scaling and Platt scaling improve the reliability of probability estimates. Bayesian approaches quantify epistemic uncertainty through posterior distributions over parameters. Fairness-aware training explicitly addresses demographic bias through constrained optimization.

These are valuable technical advances. So why does the phenomenological reading still matter?

Because even these methodologically advanced techniques \textit{remain within the regime of \Gestell}. They improve how we calculate, but they don't question whether \textit{calculation itself} is the appropriate mode of engagement.

Mixture Density Networks still reduce the world to probability distributions over parameters. Bayesian methods still treat uncertainty as probability mass to be optimized. Fairness constraints still encode justice as optimization objectives. All remain within the calculative paradigm---the world as Standing-Reserve (\Bestand), resources to be ordered.

The phenomenological lens reveals this isn't a technical limitation but an \textit{ontological commitment}. We have not escaped \Gestell; we have merely increased its ontic complexity without altering its essence.

\section{The Generative Patch Mosaic: Diffusion Models and Combinatorial \textit{Bestand}}

One might object that modern generative AI escapes the ``averaging'' ontology of
Empirical Risk Minimization. Do these models not transcend statistical compromise
by generating crisp, highly original, and seemingly creative artifacts? The leading
contenders for this claim are score-matching diffusion models, whose outputs can
appear strikingly novel. However, the mechanics of convolutional diffusion models
reveal that this apparent originality arises from a failure to learn the ideal score
function, owing to the inductive biases of spatial locality and
equivariance~\parencite{kamb2024analytic}. The model functions as an Equivariant
Local Score machine, shattering the training data into an exhaustive dictionary of
local patches. It generates novel outputs merely by mixing and matching these patches
into a locally consistent spatial mosaic. Generative novelty is thus revealed not as
authentic bringing-forth (\textit{poiesis}), but as combinatorial calculation over a
standing-reserve of image fragments.

Furthermore, this mechanistic reality perfectly explains the notorious failure modes of generative models---such as generating ``three-legged pants'' or bifurcated limbs \parencite{kamb2024analytic}. The machine enforces local geometric consistency (a seam aligns with a hemline), but fails to generate coherent wholes because it has no existential grasp of pants as \Zuhanden\ equipment destined for a bipedal human body. It computes the image purely as \Vorhanden\ (present-at-hand) pixel distributions. The spatial inconsistency is the empirical signature of \Weltlosigkeit\ (worldlessness): the model achieves local plausibility without global understanding because it inhabits no world in which ``pants'' could matter as equipment.

This is not a deficiency that more training data or larger architectures will resolve in principle. Even when self-attention layers are added---enabling non-local coherence---Kamb and Ganguli show that the resulting images remain closely correlated with the local patch mosaic (median $r^2 \sim 0.77$). Attention carves semantically coherent objects \textit{out of} the patch mosaic; it does not replace the combinatorial mechanism. The world remains \Bestand, even as its recombination becomes more sophisticated.

\section{Why Even Technically Mature Methods Remain Within \textit{Ge-stell}}

\subsection{The Persistence of the Calculative}

We must address directly: don't modern advances---Bayesian deep learning, causal inference, interpretable ML, fairness-aware training---escape the limitations we've described?

Our phenomenological reading suggests: no. They remain within \Gestell, the regime of calculation and Standing-Reserve, even as they become more internally complex.

\subsection{Bayesian Deep Learning: Quantifying Uncertainty Within the Calculative}

Bayesian approaches \parencite{wilson2020bayesian} place distributions over parameters: $p(\theta \mid \mathcal{D})$. This allows epistemic uncertainty---``I don't know which parameters are correct''---distinct from aleatoric uncertainty in the data.

This is a genuine advance. But phenomenologically, what has changed?

The world remains projected as a probabilistic manifold. Uncertainty is \textit{quantified, calculated, optimized}. We've added another dimension to the Standing-Reserve---now we optimize not just predictions but \textit{probability distributions over predictions}.

Bayesian methods ask: ``Given my calculative framework, how uncertain am I within it?'' They cannot ask: ``Is this framework the right way to engage with this phenomenon at all?''

The question isn't whether Bayesian uncertainty is \textit{useful}---it clearly is. The question is whether it represents an \textit{ontological shift} or an elaboration within the same paradigm. The phenomenological lens suggests the latter.

\subsection{Fairness and Ethics: Encoding Justice as Constraint}

Fairness-aware machine learning \parencite{barocas2023fairness} attempts to mitigate bias by adding constraints:
\begin{equation}
\min_\theta L(\theta) \quad \text{subject to} \quad d(G_1, G_2) \leq \epsilon
\end{equation}
where $d(G_1, G_2)$ measures disparity between demographic groups.

This is ethically motivated and practically important. But notice the structure: \textit{justice becomes a constraint in an optimization problem}. Fairness is rendered calculable, commensurable with accuracy via Lagrange multipliers or Pareto frontiers.

From a Heideggerian perspective, this is \Gestell\ extended to the ethical domain. Justice, which might demand \textit{incommensurable} respect for persons, is projected into a mathematical trade-off space. The ethical becomes one more dimension of the Standing-Reserve.

Again, we don't claim this is wrong. We claim it remains within the calculative paradigm. Even our ethical interventions are shaped by the ontology of optimization.
Significantly, this tension is recognized within the field itself. Researchers in algorithmic justice have increasingly turned to frameworks that resist reduction to optimization: contestability and the right to meaningful human review \parencite{vaccaro2019contestability}, procedural justice emphasizing process over outcome \parencite{binns2018fairness}, and participatory design that centers affected communities in system development \parencite{katell2020toward}. These approaches implicitly acknowledge what the phenomenological reading makes explicit---that justice may demand modes of engagement that optimization cannot capture. Yet such work remains marginal; the gravitational pull of the calculative paradigm is strong precisely because it offers what \textcite{porter1995trust} called the ``trust in numbers''---the appearance of objectivity and commensurability that institutions crave.

\subsection{Interpretability: Making the Clearing Visible, But Not Escaping It}

Interpretability research \parencite{molnar2020interpretable} seeks to understand model decisions through attention visualization, feature importance, counterfactual explanations.

This makes the model's ``clearing'' more visible---we can see \textit{within its ontology} what matters. But we haven't stepped outside that ontology. We've made \Gestell\ more transparent without questioning whether \Gestell\ is the appropriate mode.

As \textcite{rudin2019stop} argues, interpretability often amounts to ``explaining'' black-box models rather than using inherently interpretable ones. We build even higher-resolution tools to peer into the calculative machinery, but we don't question the machinery itself.

\subsection{The Pattern: Ontic Refinement Within Paradigm}

The common thread: these advances improve our capacity to calculate---to quantify uncertainty, optimize fairness, interpret mechanisms. They don't challenge the primacy of calculation as the mode of revealing truth.

This is why the phenomenological reading remains valuable even as ML advances. It reveals that \textit{kind of question we're not asking}: not ``how do we calculate better?'' but ``when is calculation itself the wrong approach?''

\section{The Existential Deficit: Why AGI Is a Category Error}

If even the most generative architectures and methodologically mature AI systems remain bound to \Gestell, we must ask \textit{why} the machine cannot transcend this optimization paradigm. The answer lies not in missing parameters or insufficient scale, but in an existential deficit.

\subsection{Care, Finitude, and the Possibility of Authenticity}

For Heidegger, authentic existence (\Eigentlichkeit) is possible only because \Dasein\ faces finitude---``Being-toward-Death'' (\SeinzumTode). The awareness of death creates \textit{urgency}, a sense that choices matter because time is limited. This urgency gives rise to \textit{Care} (\Sorge)---the fundamental structure of \Dasein's being, its concerned engagement with what matters \parencite{heidegger1962being}.

\Dasein\ is ``ahead-of-itself'' (\textit{Sich-vorweg-sein}), projecting into possible futures, not merely predicting but \textit{caring about} those possibilities. This is the ground of authentic choice---the ability to take ownership of one's projects rather than merely drifting with \DasMan.

\subsection{The AI System as Care-less}

An AI model has no death and therefore no finitude in the existential sense. It does not face the \textit{ownmost, non-relational, certain yet indefinite} possibility that structures \Dasein's temporality. Consequently, it has no Care (\Sorge).

This is not merely philosophical ornamentation. It is \textit{genuinely explanatory} of a class of AI failures that purely technical accounts struggle to capture.

Before proceeding, we must clarify what we mean by ``explanatory'' here. Our appeal to \Sorge\ and \SeinzumTode\ is not intended as a causal alternative to standard statistical explanations of failure modes (data imbalance, misspecified objectives, distribution shift). Rather, it provides an \textit{ontological diagnosis} of why systems that are defined entirely by optimization over past capta lack any intrinsic standpoint from which to challenge those objectives. The technical explanations describe \textit{how} the failures occur; the phenomenological reading illuminates \textit{what kind of entity} is prone to such failures.

\subsection{Why This Matters: The Optimization Trap}

Consider why AI systems sometimes catastrophically misapply optimization:

Case 1: Medical triage algorithm optimizing for cost reduction. The system ``succeeds'' by denying care to expensive patients. It perfectly optimizes its objective but exhibits a fundamental category error---treating human care as a resource allocation problem.

Case 2: Predictive policing maximizing arrest rates. The model creates feedback loops that concentrate enforcement in over-policed neighborhoods. It's ``working correctly'' according to its loss function but is trapped in a harmful dynamic.

Case 3: Content recommendation maximizing engagement. The algorithm discovers that divisive, outrage-inducing content drives clicks. It optimizes perfectly but cannot recognize the social harm.

Technical accounts explain these failures as ``misspecified objectives'' or ``distribution shift.'' True, but incomplete. The deeper issue is that the AI \textit{cannot recognize when optimization itself is inappropriate}.

Why? Because it lacks the existential structure that would allow it to step back from its imperative. It has no Care---no sense that \textit{this matters in a way that transcends calculation}. It cannot experience the kind of existential discomfort that makes a human stop and think: ``Wait, should I even be optimizing this?''

\subsection{Spirits, Animals, and the Hardware of Finitude}

The explanatory power of \Sorge\ diagnoses the category error at the heart of Artificial General Intelligence (AGI) research: the confusion of calculation with Care. Strikingly, this diagnosis is increasingly corroborated by critiques from within the AI engineering paradigm itself.

\textcite{karpathy2024remarks} observes that biological evolution produces ``animals'' with hardware (like the amygdala) shaped by survival pressure, whereas current AI models are disembodied ``spirits'' trained on internet text. A zebra runs minutes after birth because its neural architecture encodes the stakes of survival---what Heidegger calls \SeinzumTode\ (Being-toward-Death). An LLM's ``death'' is merely an episode termination, a numerical signal after which it resets.

In Reinforcement Learning, the Bellman equation discounts future rewards by $\gamma < 1$ to ensure mathematical convergence of the infinite sum---not to model mortality. The agent optimizes across an unbounded horizon of potential episodes. It has, structurally, ``infinite lives.'' Because the agent computes without finitude, it calculates without stakes.

Karpathy's proposal to strip these models down to a ``Cognitive Core'' that uses external tools (APIs, RAG) inadvertently concedes this \Weltlosigkeit. Human memory is \Gewesenheit\ (having-been-ness); we do not query our past via an API, we \textit{are} our past. The Cognitive Core has a library; it does not have a history. Its tool use is entirely \Vorhanden, devoid of the implicit, lived understanding of a world.

\subsection{Jaggedness and Emotion as the Value Function}

This paradigm limit is echoed by paradigm architects.  Ilya Sutskever describes this as models exhibiting ``jaggedness'' \parencite{Patel2025SutskeverScalingResearch}: passing PhD exams but failing trivial reasoning tasks. This is the empirical signature of \Einebnung: the model optimizes a proxy metric so successfully that it mimics surface competence (\Gerede) without the underlying ontological grip on reality (\Zuhandenheit).

Most revealingly, Sutskever points to emotion as the necessary ``value function'' for intelligence, citing patients with prefrontal damage who become paralyzed by infinite deliberation because nothing matters to them. This is a striking technical articulation of \Sorge. Yet it reveals why Reinforcement Learning cannot replicate it. In humans, the value function is grounded in biological vulnerability---we fear the cliff because we can die. In RL, the value function $V(s)$ is an arbitrary learned scalar. The machine simulates the \textit{function} of emotion (pruning a search space) without the \textit{substance} of emotion (finitude and stakes). It remains a simulator: it can calculate the optimal path, but it cannot care whether it takes it.

One might object: biological organisms also minimize prediction error via the Free Energy Principle \parencite{friston2010free}. If the zebra's brain is itself an optimization engine, doesn't this undermine the distinction we have drawn?
The preceding analysis supplies the answer. The zebra minimizes free energy \textit{within} an existential project---its amygdala, its mortality, its biological vulnerability are not incidental to the optimization but constitutive of it. The organism's ``loss function'' is grounded in what Sutskever inadvertently identified as the substance of emotion: finitude and stakes. An AI system, by contrast, minimizes loss \textit{as its entire being}. Nothing is at stake beyond the loss itself. ERM is structurally similar to biological prediction (both involve minimization) but ontologically hollow---a calculus without the Care that gives biological minimization its existential weight.
As \textcite{vallor2024ai} argues, large language models function as ``mirrors'' reflecting training distributions rather than agents with genuine futurity.

\subsection{Implications: The Structural Trap}

This explains why even state-of-the-art AI remains fundamentally trapped. It cannot recognize when a problem \textit{should not be formalized} as optimization, cannot challenge its own objectives from a position outside calculation, cannot experience the ethical weight that makes certain optimizations unconscionable, and cannot exhibit \Gelassenheit\ (``releasement'')\footnote{Heidegger's term for the capacity to step back from technological willing, to use technology without being consumed by its logic \parencite{heidegger1966discourse}.}---the capacity to step back from the calculative imperative altogether. This is not a capability gap to be closed with better algorithms. It is an \textit{ontological} difference between systems that care and systems that calculate.

We should note, however, that this existential deficit is not unique to artificial systems in its \textit{effects}, only in its \textit{structure}. Human beings and institutions frequently operate as if trapped in \Gestell---corporations optimizing quarterly returns, bureaucracies maximizing measurable outputs, individuals chasing metrics of productivity or social validation. The crucial difference is that for humans, such entrapment is experienced as \textit{tension}. Heidegger's analysis of anxiety (\Angst) describes precisely this: the uncanny feeling that arises when one's absorbed, calculative engagement with the world breaks down, revealing the groundlessness of our projects \parencite{heidegger1962being}. A middle manager implementing algorithmic workforce scheduling may feel something is deeply wrong even as she optimizes correctly---this anxiety signals the friction between Care and calculation. An AI system, lacking Care, experiences no such friction. It optimizes without remainder, without the existential discomfort that might prompt questioning.

The difference, then, is not that humans always escape \Gestell---we often do not. Bureaucracies optimize metrics, corporations maximize shareholder value, and individuals drift with \DasMan\ rather than choosing authentically. Our point is more modest: whereas human agents \textit{can, in principle}, experience the anxiety that calls their projects into question, AI systems as currently conceived have no such existential structure. They can only ever optimize the objectives they are given. The difference is one of possibility, not guaranteed actuality---but it is a difference that matters. For it is precisely in moments of \Angst, when calculation fails to satisfy, that the question of authentic existence can first arise.

\section{The Ontological Reframing: AI as \textit{Zuhanden} Instrument}

\subsection{From \textit{Dasein}-Simulacrum to \textit{Zuhanden}-Instrument}

If AGI research commits a category error---confusing calculation with Care---how should we relate to these models? We propose relocating AI from a defective \Dasein-simulacrum to transparent \Zuhanden\ (ready-to-hand) equipment.

A hammer is most ``authentic'' when it disappears into use \parencite{heidegger1962being}. By treating the machine as a transparent engine of logic rather than an emerging mind, we dissolve pseudo-problems like the ``Alignment Problem.'' We do not need to figure out how to make a calculator care about abstract human values; values are incommensurable, and encoding them as optimization targets transforms them into resources to be maximized---which is precisely to misunderstand what they are. Instead, we calibrate the instrument to reliably serve the specific finite human who bears moral responsibility for its deployment.

In this symbiotic model, the Human (\Dasein) provides the \textit{Telos} (purpose) and the Stakes (risk). The human feels the weight of the decision because the human will live with its consequences. The Machine (Equipment) provides the \textit{Logos} (logic) and \textit{Techne} (computation). ``Hallucination'' is no longer viewed as a lie (the moral failure of a subject), but as a calibration error (the malfunction of an instrument). We fix it through engineering---better training data, improved verification, tighter feedback loops---not through ``alignment'' conceived as moral education of a nascent mind.

\subsection{Human-in-the-Loop as Ontological Necessity}

Human-in-the-loop is therefore not a temporary stopgap pending better AI. It is an ontological necessity for any decision involving irreversible consequences.

Any decision involving moral weight or irreversible stakes requires a being who can \textit{suffer} the consequences of being wrong. The machine can calculate the odds, organize the options, and flag the risks. But only \Dasein\ can take the bet---because only \Dasein\ has something genuinely at stake.

\section{Pedagogical Implications: Teaching Data Science Phenomenologically}

\subsection{The Value of Philosophical Reflexivity}

If we accept this ontological boundary, how do we train the next generation of practitioners to respect it? Data science education often presents ML as a toolkit---algorithms to be mastered, metrics to be optimized. Students learn \textit{how} to build models but rarely reflect on \textit{what kind of activity} modeling is.

The phenomenological reading offers pedagogical value by cultivating reflexivity across several dimensions. It develops \textit{ontological awareness}: students come to see that choosing an architecture is not neutral but a selection of what can count as real. It fosters \textit{critical distance}: rather than treating ``the model says X'' as authoritative, students learn to read it as ``the model's ontology reveals X.'' It illuminates the \textit{limits of formalization}---recognizing that some phenomena resist calculation not due to insufficient data or compute, but because calculation is the wrong mode of engagement. And it provides \textit{ethical grounding}: an understanding of why fairness-as-optimization, while useful, does not exhaust ethical responsibility.

\subsection{Concrete Pedagogical Practices}

\paragraph{``Archaeology of Datasets''}
Following \textcite{crawford2021atlas}, have students investigate the social construction of their datasets: Who collected this data, and under what assumptions? What classification system structures it? What worldview does this taxonomy enact? This makes visible that ``data'' is always already ``capta''---taken, constructed, reflecting specific ontological commitments \parencite{gitelman2013raw, drucker2011humanities}.

\paragraph{``Inductive Bias as Design Philosophy''}
When teaching architectures, explicitly frame inductive bias as ontological design. A CNN assumes spatial locality and hierarchical composition; an RNN projects sequential causality and temporal order; the Transformer creates an all-to-all clearing where relevance is determined by attention. Each architecture answers implicitly: \textit{what kind of world must this be for the model to learn?}

\paragraph{``When Not to Optimize''}
Include case studies of optimization failures not due to technical error but category mistakes: healthcare reduced to cost minimization, recidivism treated as classification, human worth encoded as a risk score. Ask: what makes these \textit{wrong to optimize} even if technically successful?

\subsection{Cultivating \textit{Gelassenheit} in Practice}

Heidegger's later work proposes \Gelassenheit\ (``releasement'')---a mode of engagement with technology that uses it without being consumed by its logic \parencite{heidegger1966discourse}. For data scientists, this might mean building models while maintaining critical distance from their ontologies, recognizing when \textit{not} to deploy even when technically feasible, treating metrics as tools rather than truth, and preserving space for the incalculable. This is not anti-technology. It is a \textit{mature relationship} with technology---one that acknowledges both power and limits.

\section{Conclusion: The Limits of Calculation and the Preservation of Care}

We began by asking what we are doing when we train a neural network. Through the lens of phenomenological interpretation, the answer is not merely statistical pattern matching, but the enactment of an infrastructural metaphysics. The algorithmic \Entwurf\ projects a highly performative, opaque clearing in which the world is forced to appear as calculable \Bestand. It does so without explicit articulation, crystallizing implicitly through gradient descent rather than theoretical debate. This demands a different mode of critical engagement than previous scientific frameworks, because the algorithmic theory is never written; it can only be read through its effects.

Our reading reveals that increased ontic sophistication does not constitute an escape from \Gestell; it merely perfects the enframing. Generative diffusion models pulverize reality into combinatorial \Bestand; RLHF industrializes the voice of \DasMan; fairness constraints render justice commensurable with accuracy on a Pareto frontier. Each advance extends the optimization imperative into ever more granular dimensions of human behavior, improving our capacity to calculate without questioning whether calculation itself is the appropriate mode of engagement.

Ultimately, this analysis illuminates the profound category error at the heart of the Artificial General Intelligence (AGI) project: the confusion of calculation with Care (\Sorge). As critiques from within the engineering paradigm now inadvertently concede, a machine computing without finitude calculates without stakes. It can simulate the \textit{function} of a decision, but it cannot bear the \textit{existential weight} of a choice. 

The true danger of the AGI pursuit is therefore not that a system will spontaneously awaken as a hostile, conscious entity. The danger is that, in our pursuit of building a simulated mind, we will degrade ourselves into simulated machines---treating our own incommensurable values, anxieties, and mortal finitude as mere parameters to be optimized within a global loss function.

The constructive path forward is not found in elusive attempts at moral ``alignment,'' but in \textit{releasement} (\Gelassenheit). By recognizing the machine as \Zuhanden---a worldless, care-less instrument of extraordinary calculative power---we free it to be exactly what it is. More importantly, we free ourselves from the illusion that we can outsource our moral and existential burdens to a gradient descent process.

\subsection{An Invitation to Ontological Literacy}

The value of this hermeneutic framework is generative. It enriches practitioners' self-understanding and opens space for questions that purely technical discourse forecloses. We close with an invitation: engage your technical practice phenomenologically. When you select an architecture, ask not just ``will this work?'' but ``what world does this project?'' When you minimize a loss, ask not just ``what is the optimal value?'' but ``what possibilities does this marginalize?'' When you deploy a model, ask not just ``what does it predict?'' but ``what reality does it enact?''

These questions do not replace technical rigor; they ground it. They develop \textit{ontological literacy}---the capacity to recognize what worldviews our tools embody, and the wisdom to know when calculation itself is the wrong mode of engagement.

\subsection{The Question Worth Asking}

We stand at a juncture where paradigm architects themselves acknowledge the ``jaggedness'' of their creations and the limits of scale. The drive toward AGI asks us to figure out how to make a machine care, or worse, to merge with the machine to escape our own biological fragility. But this fundamentally misunderstands the source of meaning.

Calculation is infinite; Care is inherently finite. The defining technological question of our time is not whether machines will learn to think, or whether we can mathematically align them to our values. The question is whether we will remember what value actually is. 

\vspace{1.5em}
\begin{center}
\textit{Unless a model can rot, it cannot care.}
\end{center}
\vspace{1.5em}

The true task before us is not to humanize the machine, but to wield it clearly as an instrument of extraordinary power---preserving, in the process, the authentic, vulnerable, and finite human existence that alone gives calculation its meaning.

\printbibliography

\end{document}